\documentclass[letterpaper,twocolumn,10pt]{article}

\usepackage{usenix-2020-09}
\usepackage{amsmath}
\usepackage{enumitem}
\usepackage{xcolor}
\usepackage{graphicx}
\usepackage{subfig}
\usepackage{hyperref}

\definecolor{darkgreen}{RGB}{30, 125, 55}

\newcommand{\note}[1]{{\textcolor{blue}{*** #1 ***}}}
\newcommand{\mmio}[1]{\emph{memory-mapped I/O}}
\newcommand{\Mmio}[1]{\emph{Memory-mapped I/O}}
\newcommand{\dmap}[1]{\emph{Fastmap}}
\newcommand{\name}[1]{\emph{Talos}}
\newcommand{\master}[1]{\emph{Talos master}}
\newcommand{\primary}[1]{\emph{primary}}
\newcommand{\Primary}[1]{\emph{Primary}}
\newcommand{\primaries}[1]{\emph{primaries}}
\newcommand{\Primaries}[1]{\emph{Primaries}}
\newcommand{\backup}[1]{\emph{backup}}
\newcommand{\backups}[1]{\emph{backups}}
\newcommand{\Backup}[1]{\emph{Backup}}
\newcommand{\Backups}[1]{\emph{Backups}}
\newcommand{\server}[1]{\emph{region server}}
\newcommand{\Server}[1]{\emph{Region server}}
\newcommand{\servers}[1]{\emph{region servers}}
\newcommand{\Servers}[1]{\emph{Region servers}}
\newcommand{\replicaset}[1]{\emph{Replica Set}}
\newcommand{\comment}[1]{{}}
\newcommand{\mathvar}[1]{\text{\emph{#1}}}

\author{
  {\rm Michalis~Vardoulakis\textsuperscript{1}}\\
  Insitute of Computer Science, FORTH\\
  mvard@ics.forth.gr
  \and
  {\rm Giorgos~Saloustros}\\
  Insitute of Computer Science, FORTH\\
  gesalous@ics.forth.gr
  \and
  {\rm Pilar~Gonz\'{a}lez-F\'{e}rez}\\
  Department of Computer Engineering,\\ University of Murcia\\
  pilargf@um.es
  \and
  {\rm Angelos~Bilas\textsuperscript{1}}\\
  Insitute of Computer Science, FORTH\\
  bilas@ics.forth.gr
}

\begin{document}

\date{}

\title{Using RDMA for Efficient Index Replication in LSM Key-Value Stores}

\maketitle

\footnotetext[1]{Also with the Computer Science Department, University of Crete}

\subsection*{Abstract}

Log-Structured Merge tree (LSM tree) Key-Value (KV) stores have become a 
foundational layer in the storage stacks of datacenter and cloud services.  
Current approaches for achieving reliability and availability avoid replication 
at the KV store level and instead perform these operations at higher layers, 
e.g., the DB layer that runs on top of the KV store.  The main reason is that 
past designs for replicated KV stores favor reducing network traffic and 
increasing I/O size. Therefore, they perform costly compactions to reorganize 
data in both the primary and backup nodes.  Since all nodes in a rack-scale KV 
store function both as primary and backup nodes for different data shards 
(regions), this approach eventually hurts overall system performance.

In this paper, we design and implement \name{}, an efficient rack-scale 
LSM-based KV store that aims to significantly reduce the I/O amplification and 
CPU overhead in backup nodes and make replication in the KV store practical. We 
rely on two observations: (a) the increased use of RDMA in the datacenter, 
which reduces CPU overhead for communication, and (b) the use of KV separation 
that is becoming prevalent in modern KV stores.
We use a primary-backup replication scheme that performs compactions
only on the primary nodes and sends the pre-built index to the backup nodes of 
the region, avoiding all compactions in
backups. Our approach includes an efficient mechanism to deal with
pointer translation across nodes in the region index. Our results show
that \name{} reduces in the backup nodes, I/O amplification by up to
$3\times$, CPU overhead by up to $1.6\times$, and memory
size needed for the write path by up to $2\times$, without increasing network
bandwidth excessively, and by up to $1.3\times$. Overall, we show
that our approach has benefits even when small KV pairs dominate in a
workload (80\%-90\%). Finally, it enables KV stores to
operate with larger growth factors (from 10 to 16) to reduce space amplification
without sacrificing precious CPU cycles.

\section{Introduction}
\label{sec:intro}

Replicated persistent key-value (KV) stores are the heart of modern
datacenter storage stacks~\cite{fbookudb, cassandra,
  decandia2007dynamo, mongodb, hbase}. These systems typically use an
LSM tree~\cite{lsm} index structure
because of its 1) fast data ingestion capability for small and
variable size data items while maintaining good read and scan
performance and 2) its low space overhead on the storage
devices~\cite{rocksdbspace}. However, LSM-based KV stores suffer from
high compaction costs (I/O amplification and CPU overhead) for
reorganizing the multi-level index.

\comment{
}

To provide reliability and availability, state-of-the-art KV
stores~\cite{mongodb, cassandra} provide replication of KV pairs in
multiple, typically two or three~\cite{hdfs}, nodes. Current designs
for replication optimize network traffic and favor sequential I/O to
the storage devices both in the primary and backup
nodes. Essentially, these systems perform costly compactions to
reorganize data in both the primary and backup nodes to ensure: (a)
minimal network traffic by moving user data across nodes, and (b)
sequential device access by performing only large I/Os. However, this
approach comes at a significant increase in device traffic (I/O
amplification) and CPU utilization at the backups. Given that all
nodes in a distributed design function both as primaries and
backups at the same time, eventually this approach hurts overall
system performance. For this reason, in many cases, current approaches
for reliability and availability avoid replication at the KV store
level and instead perform these operations at higher layers, e.g. the
DB layer that runs ontop of the KV store~\cite{mongodb,cassandra}.

\comment{



}

Nowadays, state-of-the-art KV stores~\cite{mongodb, cassandra} adopt
the eager approach~\cite{rose, cassandra, mongodb} which minimizes
network traffic and recovery time at the expense of I/O amplification,
CPU, and memory overhead at the secondaries. This approach is
appropriate for systems designed for TCP/IP networks and Hard Disk
Drives (HDDs).

\comment {

}

In our work, we rely on two key observations: (a) the increased use of
RDMA in the datacenter~\cite{googlerdma,alibabardma} which reduces CPU
overhead for communication and (b) the use of KV separation that is
becoming prevalent in modern KV
stores~\cite{wisckey,diffkv,parallax,kreon,titandb,blobdb}.  Network
I/O overhead is not a significant constraint due to the use of RDMA,
especially at the rack level~\cite{Jose:2011:MDH:2066302.2066939,
  Mitchell:2013:UOR:2535461.2535475,
  Dragojevic:2014:FFR:2616448.2616486, herd}.  Also, LSM-based KV stores
tend to increasingly employ KV separation to reduce I/O
amplification~\cite{atlas,wisckey,kreon,hashkv,titandb,blobdb}.  KV
separation places KV pairs in a value log and keeps an LSM index where
values point to locations in the value log.  This technique introduces
small and random read I/Os, which fast storage devices can handle, and
reduces I/O amplification by up to 10x~\cite{vat}. Additionally,
recent work~\cite{parallax,diffkv} has significantly improved garbage
collection overhead for KV separation~\cite{hashkv, novkv} making it
production ready.

We design and implement \name{}, an efficient rack-scale
LSM-based KV store. \name{} significantly reduces I/O amplification
and CPU overhead in secondary nodes and makes replication in the KV
store practical.  \name{} uses the Kreon~\cite{kreonproject,kreon}
open-source storage engine, which is an LSM-based KV store with KV
separation, and RDMA communication for data replication, 
client-server, and server-server communication.  \name{}'s main novelty lies in the fact
that it eliminates compactions in the replicas and sends a pre-built
index from the primaries.

The three main design challenges in \name{} are the following. First,
to efficiently replicate the data (value log) \name{} uses an
efficient RDMA-based primary-backup communication protocol. This
protocol does not require the involvement of the replica CPU in
communication operations~\cite{tailwind}.

Second, since the index of the \primary{} contains pointers to its
value log, \name{} implements an efficient rewrite mechanism at the
\backups{}. More precisely, it takes advantage of the property of
Kreon that allocates its space aligned in segments(currently set to
2~MB) and creates mappings between \primary{} and \backup{}
segments. Then, it uses these mappings to rewrite pointers at the
\backups{} efficiently. This approach allows \name{} to operate at
larger growth factors to save space without significant CPU overhead.

Finally, to reduce CPU overhead for client-server communication,
\name{} implements an RDMA protocol with one-sided RDMA write
operations. \name{}'s protocol supports variable size messages that
are essential for KV stores. Since these messages are sent in a single
round trip, \name{} is able to reduce the processing overhead at the
server.

We evaluate \name{}'s performance using a modified version of the
Yahoo Cloud Service Benchmark (YCSB)~\cite{ycsb} that supports
variable key-value sizes for all YCSB workloads, similar to
Facebook's~\cite{fbook} production workloads. Our results show that
our index replication method compared to a baseline implementation
that performs compactions at the \backups{} spends $10\times$ fewer
CPU cycles per operation to replicate its index. Furthermore, it has $1.1
- 1.7\times$ higher throughput, reduces I/O amplification by $1.1 -
2.3\times,$ and increases CPU efficiency by $1.2 -
1.6\times$. Overall, \name{} technique of sending and rewriting a
pre-built index gives KV stores the ability to operate at larger
growth factors and save space without spending precious CPU
cycles~\cite{vat,rocksdbspace}.

\comment{

Scale-out KV stores~\cite{decandia2007dynamo, mongodb, hbase,rocksdb,
  leveldb} are considered as write-intensive because they typically
exhibit bursty inserts with large variations in the size of data
items~\cite{chen2012interactive, sears2012blsm}. For this reason, such
systems use at their core (a) the write-optimized LSM~\cite{lsm} data
structure to absorb bursty writes and (b) TCP-based client-server and
server-server communication for scale-out purposes. Both of these
aspects, LSM-based data organization and TCP-based communication
induce significant CPU overheads for data organization and
communication.

For this reason, there have recently been efforts to design more
lightweight approaches.  In terms of data organization, previous work
has already designed several techniques that reduce I/O amplification
in LSM-based key-value stores~\cite{all-we-know-off}. These techniques
have the effect of reducing I/O traffic (reads and writes) but also
the associated CPU overhead. In fact, with improving technology, CPU
overhead is emerging as a main bottleneck for data organization as
well~\cite{kreon,others}.  In terms of communication there is an
effort to introduce RDMA-based communication~\cite{all-we-know-off}.

Previous work~\cite{Jose:2011:MDH:2066302.2066939,
  Mitchell:2013:UOR:2535461.2535475,Dragojevic:2014:FFR:2616448.2616486,
  herd} has shown in particular, that RDMA-based protocols offer
significant gains compared to TCP/IP for in-memory key-value stores,
that incur lower overheads compared to persistent KV stores.  These
efforts are corroborated by the introduction of RDMA-based
interconnects, such as RoCE~\cite{roce,
  Dragojevic:2014:FFR:2616448.2616486} and
Infiniband~\cite{infiniband} in the
datacenter~\cite{Le:2019:ICC:3343180.3343190,Dragojevic:2014:FFR:2616448.2616486,
  Zhang:2017:PIA:3098583.3098591, Guo:2016:ROC:2934872.2934908,
  Dragojevic:2015:NCD:2815400.2815425,
  Mitchell:2013:UOR:2535461.2535475, Mittal:2015:TRC:2829988.2787510,
  Zhu:2015:CCL:2829988.2787484}.  Despite such efforts, persistent,
scale-out KV stores exhibit significant limitations.

In this paper we first identify two limitations for scale-out
persistent KV stores:

First, although using RDMA-based communication significantly reduces
overheads, it also introduces scalability challenges. RDMA requires
allocating static buffers for virtual-to-physical translation
purposes, using connection state, typically in the form of Queue
Pairs, may require synchronization across threads that issue RDMA
operations, and offers a number of different send and receive
operations. Recent work~\cite{andersen, pilaf?, etc} has focused
mainly on the selection of the most appropriate operations or protocol
optimizations to reduce the number of messages and CPU
overhead~\cite{herd, tailwind, pilaf?, more}. However, most approaches
used are not able to scale with the number of clients or the number of
key regions.

Second, replication of LSM-based structures across servers for
reliability purposes is still an expensive operation, even when using
RDMA-based communication. While using RDMA results in spending less
CPU cycles on networking, servers will still have to perform data
reorganization in order for failure recovery to be practical, which is
the where most CPU cycles are spent in an LSM-based key-value store.
\note{ab: discuss in detail what the problem is and where the
  overheads come from.}

In this paper we design, \name{}, a scale-out persistent KV-store that
addresses these two issues. \name{} is a rack scale key-value store
that uses fast storage devices and RDMA-based networking to increase
CPU efficiency.  At its core \name{} uses Kreon, to store data on
local flash storage devices. Kreon is an efficient key-value store
which uses key-value separation (KV) and organizes its levels in a
fine-grained manner with B+tree indexing to increase CPU efficiency.

In this work we design a protocol for RDMA-based communication that is
able to scale in all directions: number of clients, number of key
regions, and number of servers. In addition, we look at more intricate
issues of RDMA-based communication, namely spinning for arriving
messages. Low-latency RDMA-based protocols use spinning instead of
sleeping to reduce latency for message detection. However, spinning is
not an acceptable operation for modern datacenters that are limited by
power. In our work, we design an adaptive scheme where a single thread
in each server performs adaptive spinning/sleeping, while all other
worker threads only wake-up when necessary. Our approach avoids
unnecessary spinning and at the same time is able to quickly react to
network requests.  \note{ab: need to be more specific and clear here
  about what we do in this paper. Also, organize better in terms of
  what we do, how we do it, what is the impact.}  \note{mvard: We have
  one spinning thread per 4 workers, which never sleeps. The worker
  threads do indeed sleep while they have no work to do.}

KV stores~\cite{decandia2007dynamo, mongodb, hbase,rocksdb, leveldb}
usually use TCP/IP since this protocol is widely deployed and the
majority of datacenter infrastructure supports it.  However nowadays,
TCP/IP fails to satisfy applications increasing demand for higher
throughput and lower latency. stack requiring extensive computing
power. Consequently it inherently incurs high overheads due to its
streaming semantics, and it leaves little processing resources for
applications~\cite{Bierbaum:2002:MET:648047.745852}.  In addition,
even though during the last decade Ethernet technology has evolved to
provide link speeds up to of 400 Gbps, TCP/IP protocol fails to
deliver the expected performance due its high latency and CPU
overhead~\cite{Marinos:2014:NSS:2740070.2626311,pilar2014,mTCP}.
TCP/IP stack cannot provide the higher bandwidth and lower latency
that datacenter networks should offer, limiting the extent to which
applications can make use of these characteristics.

Then, we design a primary-backup replication strategy over
RDMA~\cite{primary-backup,chainrepl} that dramatically reduces
replication overhead in the secondary by taking advantage of the work
already performed in the primary to index data. All index updates are
performed on new tree nodes instead of in place. This way a master can
easily send to its replicas the changes applied to a level's index
after a data reorganization operation. Moreover, since allocations on
the underlying storage device are performed in fixed size segments,
the only extra information needed by a replica in order to reuse the
primary's index is a map of the primary's segments to the replica's
corresponding segments. By sending the index, servers don't have to
perform costly data reorganization operations for replica data,
meaning increased CPU efficiency and lower I/O amplification at the
expense of increased network traffic.  \note{ab: describe in 1
  paragraph how we move indexes, logs, pointers and why this reduces
  overheads.}  \note{ab: describe in 2 sentences how we rebuild the
  secondary and primary after a failure.}

\note{ab: Giorgo, describe here the tradeoffs you have been mentioning
  about CPU vs. network and what we do (in contrast to what is
  happening today).  make the discussion specific to kv stores and
  what/how they send data/index, what is the data volume, etc.}

The rest of this paper is organized as follows. \note{ab: fill in}

}

\section{Background}
\label{sec:background}

\paragraph{LSM tree}
LSM tree~\cite{lsm} is a write-optimized data structure that organizes
its data in multiple hierarchical levels. These levels grow
exponentially by a constant growth factor $f$.  The first level
($L_0$) is kept in memory, whereas the rest of the levels are on the
storage device.  \comment{ During an insert operation, LSM tree stores
  the key-value pair in the $L_0$ memory component. When the size of a
  level $L_i$ exceeds its maximum capacity, it merges-sorts $L_i$ with
  $L_{i+1}$ data into a new $L'_{i+1}$ to free up space in upper
  levels. This process is called {\em compaction}, and it is the core
  mechanism through which the LSM tree achieves 1) high ingestion
  rates and 2) efficient space management. However, compaction
  produces excess read and write I/O traffic named {\em
    amplification}.  } There are different ways to organize data
across LSM levels~\cite{tiering,lsm}. However, in this work, we focus
on leveled LSM KV stores that organize each level in non-overlapping
ranges.

In LSM KV stores, the growth factor $f$ determines how much the levels
grow. The minimum I/O amplification is for $f=4$~\cite{vat}.  However,
systems choose larger growth factors (typically 8-10) because a larger
$f$ reduces space amplification at the expense of increased I/O
amplification for high update ratios, assuming that intermediate
levels contain only update and delete operations.  For example,
assuming an $L_0$ size of 64~MB and a device of 4~TB capacity an
increase of the growth factor from 10 to 16 results in 6\% space
savings.

\paragraph{KV separation}
Current KV store designs~\cite{atlas,wisckey,kreon,hashkv,blobdb} use
the ability of fast storage devices to operate at a high (close to
80\%~\cite{kreon}) percentage of their maximum read throughput under
small and random I/Os to reduce I/O amplification. The main techniques
are KV separation~\cite{atlas,wisckey,kreon,hashkv,blobdb,titandb} and
hybrid KV placement~\cite{parallax,diffkv}.  KV separation appends
values in a value log instead of storing values with the keys in the
index. As a result, they only re-organize the keys (and pointers) in
the multi-level structure, which, depending on the KV pair sizes, can
reduce I/O amplification by up to 10x~\cite{vat}.  Hybrid KV
placement~\cite{diffkv,parallax} is a technique that extends KV
separation and solves problem of the garbage collection overhead introduced by 
the garbage collection in the value log, especially for
$medium\geq 100~B$ and $ large\geq 1000~B$ KV pairs.

\paragraph{RDMA}
RDMA verbs is a low-level API for RDMA-enabled applications. The verbs
API operates atop of various lossless transport protocols such as
Infiniband or RDMA over Converged Ethernet (RoCE).  The verbs API supports
\emph{two-sided} send/receive message operations and \emph{one-sided} RDMA 
read/write operations. In send
and receive operations, both the sender and the receiver actively
participate in the communication, consuming CPU cycles.  RDMA read and
write operations allow one peer to directly read or write the memory
of a remote peer without the remote one having to post an operation,
hence bypassing the remote node CPU and consuming CPU cycles only in
the originating node.


\paragraph{Kreon}
\label{subsec:kreon}

Kreon~\cite{kreon, kreonproject} is an open-source persistent LSM-based KV
store designed for fast storage devices (NVMe).  Kreon increases CPU
efficiency and reduces I/O amplification using (a) KV separation, and
(b) \Mmio{} for its I/O cache and direct I/O for writing data to the
device.  To perform KV separation~\cite{wisckey,tucana,atlas,jungle},
Kreon stores key-value pairs in a log and keeps an index with pointers
to the key-value log. This technique reduces I/O amplification up to
$10\times$~\cite{vat}.  A multilevel LSM structure is used to organize its
index.  The first level $L_0$ lies in memory, whereas the rest of the levels
are on the device.  Kreon organizes each level internally as a B+-
tree where its leaves contain the <key\_prefix, value\_location>
pairs.  Finally, all logical structures (level's indexes and value
log) are represented as a list of segments on the device. The segment
is the basic allocation unit in Kreon and is currently set to
2~MB. All segment allocations in Kreon are segment aligned.

Kreon uses two different I/O paths: \mmio{} to manage its I/O cache, access the 
storage devices during read and scan operations, and to write its value log.  
Furthermore, it uses direct I/O to read and write the levels during 
compactions.

\section{Design}

\subsection{Overview}
\label{sec:overview}
\name{} is a persistent rack-scale KV store that increases CPU
efficiency in \backup{} regions for data replication purposes.
\name{} uses a primary-backup protocol~\cite{primarybackup} for
replicating the data via RDMA writes without involving the CPU of the
\backups{}~\cite{tailwind} for efficiency purposes.  To reduce the
overhead of keeping an up-to-date index at the \backups{}, we design
and implement the \emph{Send Index} operation for systems that use
KV-separation~\cite{wisckey,kreon,titandb,blobdb} or hybrid KV
placement~\cite{parallax,diffkv}.  \Primary{} servers, after
performing a compaction from level $L_i$ to $L_{i+1}$, send the
resulting $L'_{i+1}$ to their \backups{} in order to eliminate
compactions in \backup{} regions.  Because $L'_{i+1}$ contains
references to the primary's storage address space, \backups{} use a
lightweight rewrite mechanism to convert the \primary{}'s $L'_{i+1}$
into a valid index for their own storage address space. During the
Send Index operation, the \backup{} uses metadata (hundreds of~KB)
retrieved during the replication of the KV log to translate pointers
of the \primary{}'s KV log into its own storage space.

We design and implement an RDMA Write-based protocol for both its
server-server and client-server communication. We build our protocol
using one-sided RDMA write operations because they reduce the network
processing CPU overhead at the server~\cite{andersen} due to the
absence of network interrupts. Furthermore, \name{}, as a KV store,
must support variable size messages. We design our protocol to
complete all KV operations in a single round trip to reduce the
messages processed per operation by the servers.


\name{} uses Kreon~\cite{kreonproject,kreon} KV store for efficiently
managing the index over its local devices. We modify Kreon to use
direct I/O to write its KV log to further reduce CPU overhead for
consecutive write page faults, since write I/Os are always
large. Direct I/O also avoids polluting the buffer cache from
compaction traffic.

\begin{figure}
\centering
\includegraphics[width=0.55\columnwidth]{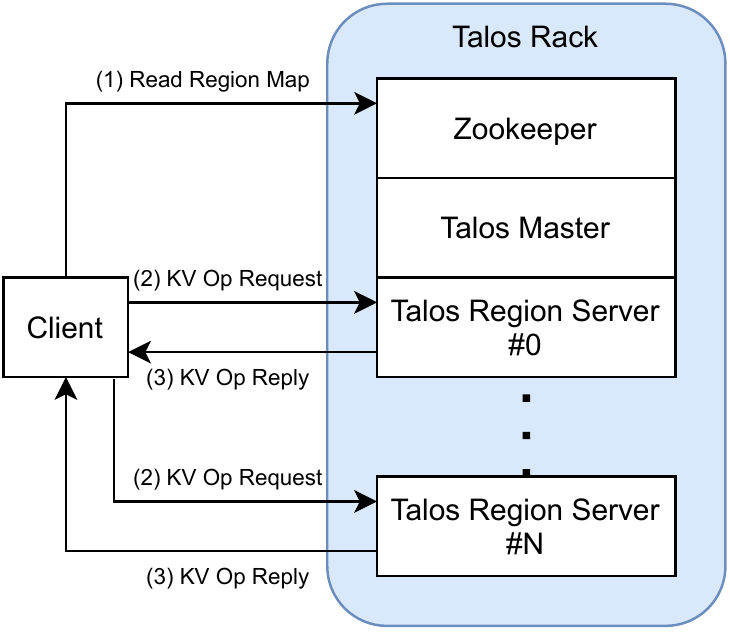}
\caption{\name{} overview.}
\label{fig:tebis}
\end{figure}

Finally, \name{} partitions the key-value space into non-overlapping
key ranges named \emph{regions} and offers clients a CRUD API (Create,
Read, Update, Delete) as well as range (scan) queries.  \name{}
consists of the three following entities, as shown in
Figure~\ref{fig:tebis}:

\begin{enumerate}[topsep=2pt]
\item \emph{Zookeeper}~\cite{zookeeper}, a highly available service
  that keeps the metadata of \name{} highly available and strongly
  consistent, and checks for the health of \name{} \servers{}.
\item \Servers{}, which keep a subset of regions for which they either
  have the \primary{} or the \backup{} role.
\item \master{}, which is responsible for assigning regions to
  \servers{} and orchestrating recovery operations after a failure.
\end{enumerate}

\subsection{Primary-backup Value Log Replication}
\label{sec:replication}

\begin{figure*}[t]
  \subfloat[][Key-Value log replication %
    process]{\includegraphics[width=\textwidth]{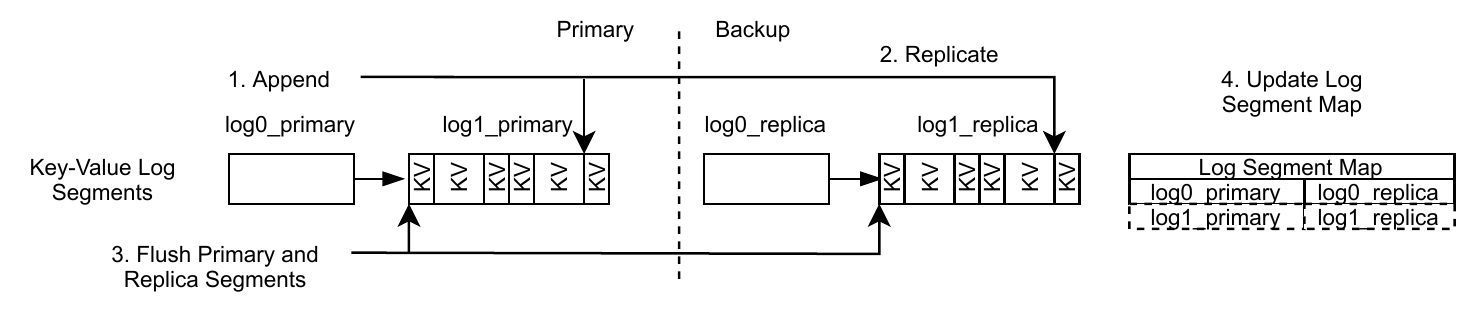}
    \label{fig:log_replication}
  }

  \subfloat[][Key-Value index replication %
    process]{\includegraphics[width=\textwidth]{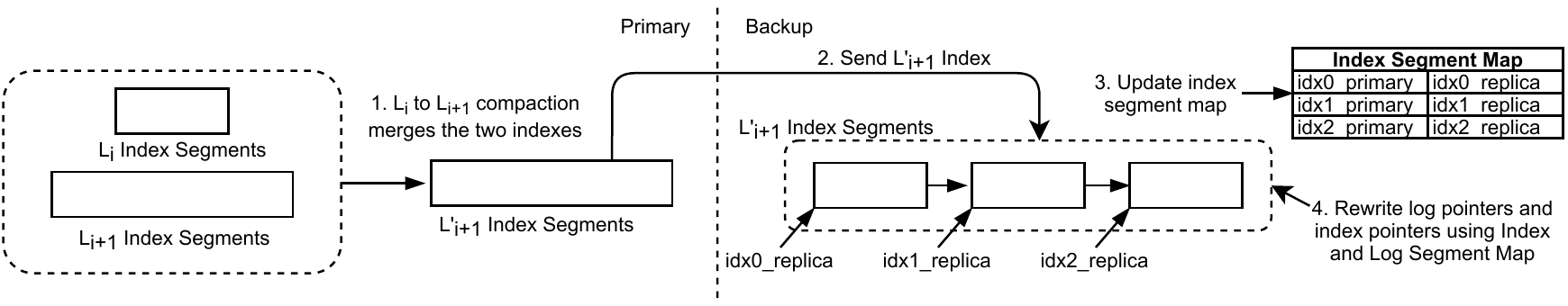}
    \label{fig:index_replication}
  }
  \caption{Replication in \name{}.}
\end{figure*}

We design and implement a primary-backup replication protocol to
remain available and avoid data loss in case of failures. Each \server{} stores 
a set of regions and has either the \primary{} or \backup{} role for any region 
in its set. The main design challenge that \name{} addresses is to replicate 
data and keep full indexes at the \backups{} with low CPU overhead. Having an
up-to-date index at each \backup{} is necessary to provide a fast recovery time 
in case of a failure.

\name{} implements a primary-backup protocol over RDMA for
replication~\cite{primarybackup,tailwind}.  On initialization, the
\primary{} sends a message to each \backup{} to request an RDMA buffer
of the same size as Kreon's value log segment (2~MB). When a client
issues an insert or update KV operation, the \primary{} replicates
this operation in its set of \backup{} servers. The \primary{}
completes the client's operation in the following three steps:
\begin{enumerate}[topsep=2pt]
  \item Inserts the KV pair in Kreon, which returns the offset of the
    KV pair in the value log, as shown in step 1 in
    Figure~\ref{fig:log_replication}).
  \item Appends (via RDMA write operation) the KV pair to the RDMA
    buffer of each replica at the corresponding offset, as shown in
    step 2 in Figure~\ref{fig:log_replication}).
  \item Sends a reply to the client after receiving the completion
    event from all \backup{}s for the above RDMA write operation.
\end{enumerate}
The \backup{}'s CPU is not involved in any of the above steps due to
the use of RDMA write operations. When a client receives an
acknowledgment it means that its operations has been replicated to all
the memories of the replica set.

When the last log segment of the \primary{} becomes full, the
\primary{} writes this log segment to persistent storage and sends a
\emph{flush} message to each \backup{} requesting them to persist
their RDMA buffer, as shown in step 3 in
Figure~\ref{fig:log_replication}. \Backup{} servers then copy their
RDMA buffer to the last log segment of the corresponding Kreon region
and write that log segment to their persistent storage. \Backup{}
servers also update their \emph{log segment map}, as shown in step 4
in Figure~\ref{fig:log_replication}. The log segment map contains entries
of the form <primary value log segment, replica value log segment>.
Each \backup{} server keeps this map and updates it after each flush
message. \Backups{} use this map to rewrite the \primary{} index. We will 
discuss this index rewriting mechanism in more detail in 
Section~\ref{subsubsec:index}.

Each \backup{} keeps the log segment map per \backup{} region in
memory. The log segment map has a small memory footprint; for a 1~TB
device capacity and two replicas, the value log will be 512~GB in the worst 
case. With the segment size set to 2~MB, the memory size of the log
segment map across all regions will be at most 4~MB. In case of
\primary{} failure, the new \primary{} informs its \backups{} about
the new mappings.

\subsection{Efficient \Backup{} Index Construction}
\label{subsubsec:index}

\name{} instead of repeating the compaction process at each server to
reduce network traffic, takes a radical approach. \Primary{} executes
the heavy, in terms of CPU, compaction process of $L_i$ and $L_{i+1}$
and sends the resulting $L'_{i+1}$ to the \backups{}. This approach
has the following advantages. First, servers do not need to keep an
$L_0$ level for their \backup{} regions, reducing the memory
budget for $L_0$ by $2\times$ when keeping one replica per region or by 
$3\times$ when keeping two replicas.  Second, \backups{} save device
I/O and CPU since they do not perform compactions for their \backup{}
regions.

Essentially, this approach trades network I/O traffic for CPU, device
I/O, and memory at servers since network bulk transfers are CPU
efficient due to RDMA.  The main design challenge to address is
sending the \primary{} level index to the \backup{} in a format that
\backups{} can rewrite with low CPU overhead. \name{} implements the
rewriting process at the \backup{} as follows.

When level $L_i$ of a region in a \primary{} is full, \name{} starts a
compaction process to compact $L_i$ with $L_{i+1}$ into
$L'_{i+1}$. The \primary{} reads $L_i$ and $L_{i+1}$ and builds
$L'_{i+1}$ B+-tree index.  $L'_{i+1}$ is represented in the device as
a list of segments (currently set to 2~MB) which contains either leaf
or index nodes of the B+-tree. Leaf nodes contain pairs of the form
<key prefix, pointer to value log> whereas index nodes contains pairs
of the form <pivot, pointer to node>.

To transfer the $L'_{i+1}$ index, the \primary{} initially requests
from each \backup{} to register an RDMA buffer of segment size.
\name{} only uses these buffers during the $L'_{i+1}$ index transfer and 
deregisters and frees them once the compaction is completed.


On receiving a leaf segment, each \backup{} \server{} parses it and
performs the following steps. Leaf segments contain key prefixes that
work across both indexes. The \backup{} has to rewrite pointers to the
value log before using them.  \name{}'s storage engine performs all
allocations in 2~MB aligned segments. As a result, the first high 22
bits of a device offset refer to the segment's start device offset.
The remaining bits are an offset within that segment. To rewrite the
value log pointers of the \primary{}, the \backup{} first calculates the 
segment start offset of each KV pair. Since all segments are aligned, it does 
this through a modulo
operation with segment size. Then it issues a lookup in the log map
and replaces the primary segment address with its local segment
address.

For index segments, \name{} keeps in-memory an \emph{index map} for
the duration of $L'_{i+1}$ compaction.  \Backups{} add entries to this
map whenever they receive an index segment from the \primary{}. This map
contains entries using as the \primary{}'s index segment as the key and the 
corresponding \backup{}'s index segment as the value, as shown in
Figure~\ref{fig:index_replication}. This mechanism translates pointers
to index or leaf nodes within the segment the same way as it does for
value log pointers in leaves.  Finally, on compaction completion,
the \primary{} sends the root node offset of $L_{i+1}$ to each \backup{},
which each \backup{} translates to its storage space.

\subsection{Failure Detection}
\label {sec:failure_detection}

\name{} uses Zookeeper's ephemeral nodes to detect failures. An
ephemeral node is a node in Zookeeper that gets automatically deleted
when its creator stops responding to heartbeats of Zookeeper. Every
\server{} creates an ephemeral node during its initialization.  In
case of a failure, the \master{} gets notified about the failure and
runs the corresponding recovery operation.  In case of \master{}
failure, all \servers{} get notified about its failure through the
ephemeral node mechanism. Then, they run an election process through
Zookeeper and decide which node takes over as \master{}.

\subsection{Failure Recovery}

\name{} uses Zookeeper, similar to other systems~\cite{hbase,kudu}, to
store its \emph{region map}. Each entry in the region map consists of
the range of the region <start key, end key>, the \primary{} server
responsible for it, and the list of its \backup{} servers.  The region
map is infrequently updated when a region is created, deleted after a
failure, or during load-balancing operations.  Therefore, in \name{}
Zookeeper operations are not in the common path of data access.

The \master{} reads the region map during its initialization and
issues \emph{open region} commands to each \server{} in the \name{}
cluster, assigning them a \primary{} or a \backup{} role. After
initialization, the role of the \master{} is to orchestrate the
recovery process in case of failures and to perform load balancing
operations.

Clients read and cache the region map during their
initialization. Before each KV operation, clients look up their local
copy of the region map to determine the \primary{} \server{} where
they should send their request. Clients cache the region map since
each region entry is 64~B, meaning just 640~KB are enough for a region
map with 10,000 regions, and changes to it are infrequent.  When a
client issues a KV operation to a \server{} that is not currently
responsible for the corresponding range, the \server{} instructs it to
update their region map.

\name{} has to handle three distinct failure cases: 1) \backup{}
failure, 2) \primary{} failure, and 3) \master{} failure. Since each
\name{} \server{} is part of multiple region groups, a single node
failure results in numerous \primary{} and \backup{} failures, which
the \master{} handles concurrently.  First, we discuss how we handle
\backup{} failures.

In case of a \backup{} failure, the \master{} replaces the crashed
\server{} with another one that is not already part of the
corresponding region's group. The \master{} then instructs the rest of
the \servers{} in the group to transfer the region data to the new
member of the region group.  The region experiencing the \backup{}
failure will remain available throughout the whole process since its
\primary{} is unaffected. However, during the reconstruction of the
new \backup{}, the region data are more susceptible to future
failures, since there's one less \backup{} copy.

In case of a \primary{} failure, the \master{} first promotes one of
the existing \backup{} \servers{} in that region group to the
\primary{} role, and updates the region map.  The new \primary{}
already has a complete KV log and an index for levels $L_{i}$, where
$i \geq 1$. The new \primary{} \server{} replays the last few segments
of its value log in order to construct an $L_0$ index in its memory
before being able to server client requests. Now that a new \primary{}
\server{} exists for the group, the \master{} handles this failure as
if it were a \backup{} failure.  During the \primary{} reconstruction
process, \name{} cannot server client requests from the affected
region.

When the \master{} crashes, the rest of the \servers{} in the \name{}
cluster will be notified through Zookeeper's ephemeral node mechanism,
as discussed in Section \ref{sec:failure_detection}. They will then
use Zookeeper in order to elect a new \master{}. During the \master{}
downtime, \name{} cannot handle any region failures, meaning that any
region that has suffered a \primary{} failure will remain unavailable
until a new \master{} is elected and initiates the recovery process
for any regions that suffered a failure.

\subsection{RDMA Write-based Communication Protocol}




\begin{figure}
  \centering
  \includegraphics[width=0.7\columnwidth]{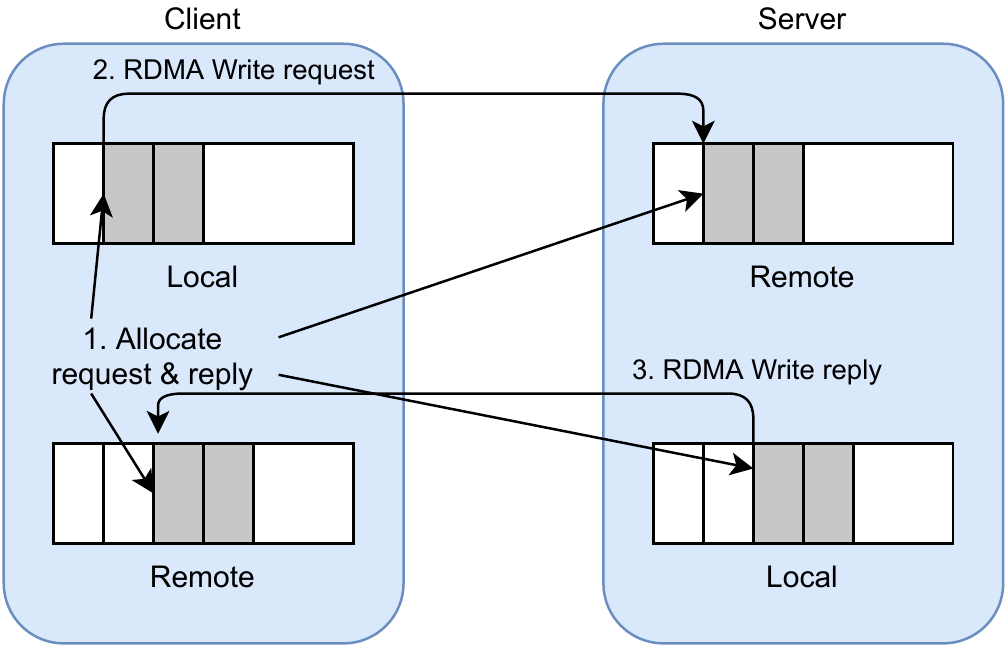}
  \caption{Allocation and request-reply flow of our RDMA Write-based
    communication protocol.}
  \label{fig:circular-buffers}
\end{figure}

\name{} performs all client server communication via one-sided RDMA
write operations~\cite{andersen} to avoid network interrupts and thus
reduce the CPU overhead in the server's receive
path~\cite{andersen,herd}.  Furthermore, to avoid the overhead of
registering and deregistering RDMA memory buffers per KV operation,
the server and client allocate a pair of buffers during queue pair
(QP) creation.  Their size is dynamic within a range (256~KB
currently) set by the client on QP creation. \server{} frees this
buffer when a client disconnects or suffers a failure. The client
manages these buffers to improve CPU efficiency in the server.
 
Clients allocate a pair of messages for each KV operation; one for
their request and one for the server's reply.  All buffers sizes are
multiples of a size unit named \emph{message segment size} (currently
set to 128 bytes).  Clients put in the header of each request the
offset at their remote buffer where \server{} can write its reply.
Upon completion of a request, the \server{} prepares the request's
reply in the corresponding \emph{local} circular buffer at the offset
supplied by the client.  Then it issues an RDMA write operation to the
client's \emph{remote} circular buffer at the exact
offset. Figure~\ref{fig:circular-buffers} shows a visual
representation of these steps.  As a result, the \server{} avoids
expensive synchronization operations between its workers to allocate
space in the remote client buffers and update buffer state metadata
(free or reserved). If the client allocates a reply of insufficient
size, the \server{} sends part of the reply and informs the client to
retrieve the rest of the data.

\subsubsection{Receive Path}

\begin{figure}
  \includegraphics[width=\linewidth]{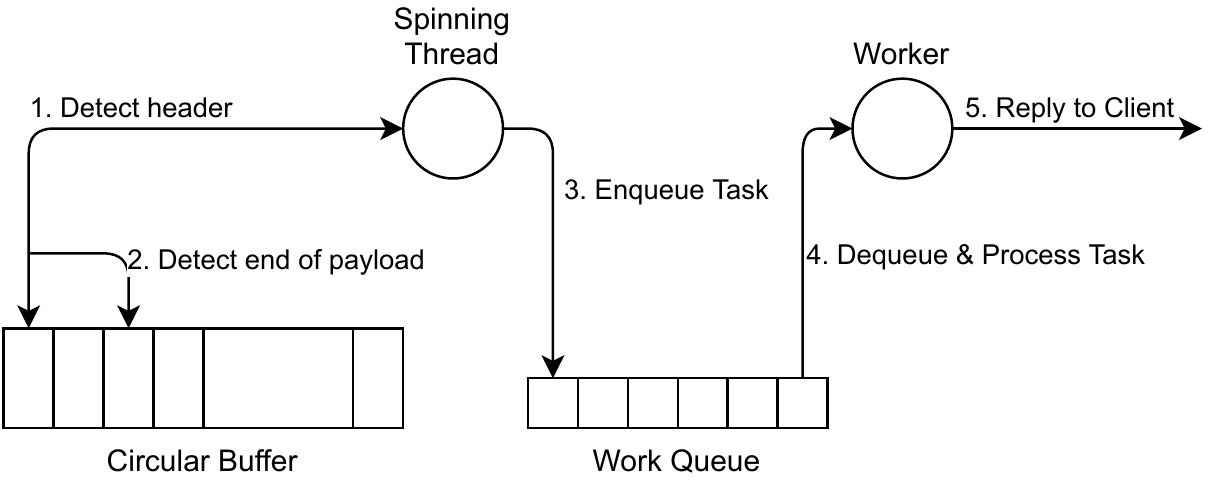}
  \caption{Message detection and task processing pipeline in
    \name{}. For simplicity, we only draw one circular buffer and a
    single worker.}
  \label{fig:tebis_arch}
\end{figure}

To detect incoming messages, in the absence of network interrupts,
each \server{} has a \emph{spinning thread} which spins on predefined
memory locations in its corresponding clients' remote circular
buffers, named \emph{rendezvous points}.  The spinning thread detects
a new message by checking for a magic number in the last field of the
message header, called the \emph{receive field}, at the next
rendezvous location. After it detects a new message header, it reads
the payload size from the message header to determine the location of
the message's tail.  Upon successful arrival of the tail, it assigns
the new client request to one of its workers and advances to the next
rendezvous location of the circular buffer.

To support variable size messages \name{} adds appropriate padding so
that their size is message segment aligned. This quantization has two
benefits: 1) Possible rendezvous points are at the start of each
segment, offset by the size of a message header minus the size of the
header's receive field.  Upon reception of a message the \server{}
advances its rendezvous point by adding the current message size.  2)
The \server{} doesn't have to zero the whole message upon each request
completion; it only zeros the possible rendezvous points in the
message segments where the request was written.

\subsubsection{Reset Operation in Circular Buffers}
There are two ways to reset the rendezvous point to the start of the
circular buffer: 1) When the last message received in the circular
buffer takes up its whole space, the server will pick the start of the
circular buffer as the next rendezvous location, and 2) When the
remaining space in the circular buffer is not enough for the client to
send their next message, they will have to circle back to the start of
the buffer. In this case, they will send a \emph{reset rendezvous}
message to inform the server that the next rendezvous location is now
at the start of the circular buffer.

\subsubsection{Task Scheduling} To limit the max number of threads, \name{}
uses a configurable number of workers. Each worker has a private
\emph{task queue} to avoid CPU overhead associated with contention on
shared data. In this queue, the spinning thread places new tasks, as
shown in Figure~\ref{fig:tebis_arch}. Workers poll their queue to
retrieve a new request and sleep if no new task is retrieved within a
set period of time (currently $100~\mu s$).  The primary goal of
\name{}'s task scheduling policy is to limit the number of wake-up
operations, since they include crossings between user and kernel
space.  The spinning thread assigns a new task to the current worker
unless its task queue has more than a set amount of tasks already
enqueued. In the latter case, the spinning thread selects a running
worker with less than that set amount of queued tasks and assigns to
it the new task. If all running workers already exceed the task queue
limit, the spinning thread wakes up a sleeping worker and enqueues
this task to their task queue.

\comment{
}

\comment{

}

\comment{

}

\comment{
}

\comment{
\subsection{Connection State}

A connection in \name{} consists of the RDMA QP and the two circular
buffers used to prepare and receive messages. In order for a client to
send a request to a \server{}, the client allocates enough space in
their send circular buffer to prepare their request, as well as enough
space in their receive circular buffer, where the \server{} will write
their reply to the client's request. The offset and length of the
allocated reply is added in the message's header to be used later by
the \server{} The client will then send the message to the
corresponding \server{} by writing through RDMA to the same offset of
the \server{}'s receive circular buffer.  The \server{} will allocate
their reply in their send circular buffer at the offset written by the
client in the message's header.  The reply is then sent to the
client's receive circular buffer using RDMA write.  The client will
free the allocated space for the request and its reply once it
receives the \server{}'s reply.

\note{ab: need a figure with circular buffers between two nodes, other
  connection elements, how messages are placed, detected, etc.}

When issuing a request, clients use the region map to identify the
primary for a given key and any existing connections to that
server. If the client does not have an open connection with the
corresponding server, they will open one using the RDMA Communication
Manager library. All further communication is through our custom
messaging protocol built on top of RDMA write. They exchange the
necessary information to perform RDMA writes using RDMA send \&
receive operations.

Each \server{} maintains a linked list of all incoming client
connections.  Each connection has a \emph{rendezvous} point where the
next message header will arrive.  We support variable size messages by
splitting the circular buffers into fixed size chunks so that
rendezvous points are at fixed locations. In this manner, we only need
to zero those fixed locations in order to be able to detect new
messages instead of having to zero the entire circular buffer.  We pad
messages when necessary so that their total size is a multiple of the
chunk size.  The rendezvous location is initialized when a new
connection is accepted, and then advances to the next location, which
is calculated by the chunk size of the last message.  \note{Michalis:
  when is the rendezvous location reset?}  In order to detect new
incoming requests, each \server{} creates a spinning thread that
iterates over their connection list, checking each client connection's
rendezvous points. \note{Michalis: Should we say a spinning thread per
  numa node?}  }

\comment{
\subsection{Scalability with number of connections (clients and servers)}

\note{Connection state for millions of connections}

\subsection{Elasticity with the number of clients}

\note{Buffer management to achieve ALWAYS max network throughput}

\subsection{Scalability with number of intra-node threads}

\note{synchronization for accessing buffers}

\note{synchronization for accessing connections}

\note{scanning for arriving messages}

\note{spinning waiting for arriving messages}

\note{waking up workers}

\note{workers going back to sleep}

\note{master threads going to sleep}

\note{ab: explain here a bit more what are the challenges. discuss QP,
  card state, buffers, threads, receive side spinning, possibly RDMA
  message types.}



\note{ab: structure the rest to answer the challenges above.}

\note{Michalis: we need to mention something about the recovery
  process as well as describe a plausible design for leader election.}

When a \primary{}, \backup{} failure is detected, Zookeeper triggers
the corresponding recovery operation.

\note{ab: how many and what recovery operations do we have?} We note
that \name{} does not do anything about client failures, as they are
not a necessary part of the recovery scheme.

Upon a server (\primary{} or \backup{}) failure, Zookeeper will
trigger the recovery process for all regions hosted by the node, both
with \primary{} and \backup{} role. When the recovery process is
complete, Zookeeper will have an updated region map.

\note{fill in with our protocol for recreating primaries and
  \backup{}s for each region when a node fails.}

\subsubsection{Zero Client Expectations:}
Clients do not participate in any way in system recovery process. They
merely send requests and receive responses, while they detect server
failures via timeouts and heartbeats.

Clients initially communicate with Zookeeper to retrieve a
\emph{region map} of the system that lists the \primary{} and
\backup{} server per region. Regions in \name{} are coarse grain and
the region map is typically a few MBytes for datasets in the order of
100s of TBytes. For instance, assuming that each region is a few 10s
of GBytes, then each server can host a few thousand regions (10s of
TBytes), and a system with several 10s of servers can host PBytes of
data. Cummulatively, a few hundred thousand of regions require a few
MBytes to describe.

Then, clients can start issuing operations (Put, Get, Delete, Scan) to
any region they desire. Operations are always issued to the
\primary{}.  While a client is waiting for a reply, they will send a
heartbeat to the corresponding \server{} by posting a zero-length RDMA
write to that \server{}. If that operation completes successfully, the
client will then keep waiting for the reply.

If a client is not able to contact a \primary{} or does not receive a
response for a pending request, it assumes that its region map is
stale and will refresh this information from Zookeeper.  When servers
receive client requests, they only respond to requests for which they
are primaries. If they do not handle the requested region or they
serve as a \backup{}, then they merely drop the request. This will
result in the client timeout and refreshing its region map from
Zookeeper. After a client receives the new region map, it merely
retries its operation with the new region \primary{}.



}

\section{Evaluation Methodology}
\label{sec:platform}

Our testbed consists of two servers where we run the KV store.  The
servers are identical and are equipped with an AMD EPYC 7551P
processor running at 2~GHz with 32 cores and 128~GB of DDR3 DRAM. Each
server uses as a storage device a 1.5~TB Samsung NVMe from the PM173X
series and a Mellanox ConnectX 3 Pro RDMA network card with a
bandwidth of 56~Gbps. We limit the buffer cache used by \name{}'s
storage engine (Kreon) using \emph{cgroups} to be a quarter of the
dataset in all cases.

In our experiments, we run the YCSB benchmark~\cite{ycsb} workloads
Load A and Run A -- Run D. Table~\ref{tab:ycsbworkloads} summarizes
the operations run during each workload.  We use a C++ version of
YCSB~\cite{Jinglei2016} and we modify it to produce different values
according to the KV pair size distribution we study. We run \name{}
with a total of 32 regions across both servers. Each server serves as
\primary{} for the 16 and as \backup{} for the other 16. Furthermore,
each server has 2 spinning threads and 8 worker threads in all
experiments. The remaining cores in the server are used for
compactions.

\begin{table}
  \centering
  \begin{tabular}{lc}
    & \textbf{Workload}\\ \hline Load A & 100\% inserts \\ Run A &
    50\% reads, 50\% updates \\ Run B & 95\% reads, 5\% updates \\ Run
    C & 100\% reads \\ Run D & 95\% reads, 5\% inserts \\ \hline
  \end{tabular}
  \caption{\label{tab:ycsbworkloads} Workloads evaluated with
    YCSB. All workloads use a Zipfian distribution except for Run D
    that use latest distribution.}
\end{table}

In our evaluation, we also vary the KV pair sizes according to the KV
sizes proposed by Facebook~\cite{fbook}, as shown in
Table~\ref{tab:kv-dist}.  We first evaluate the following workloads
where all KV pairs have the same size: Small (S), Medium (M), and
Large (L).

Then, we evaluate workloads that use mixes of S, M, and L KV pairs.
We use small-dominated (SD) KV size distribution proposed by
Facebook~\cite{fbook}, as well as two new mixed workloads: \emph{MD}
(medium dominated) and \emph{LD} (large dominated). We summarize these
KV size distributions in Table~\ref{tab:kv-dist}.

\begin{table}
  \centering
  \begin{tabular}{lcccc}
    & KV Size Mix &            & Cache per  & Dataset\\
    & S\%-M\%-L\% & \#KV Pairs & Server (GB) & Size (GB)\\
    \hline
      S  & 100-0-0  & 100M & 0.38 & 3\\
      M  & 0-100-0  & 100M & 1.4 & 11.4\\
      L  & 0-0-100  & 100M & 11.9 & 95.2\\
      SD & 60-20-20 & 100M & 2.8 & 23.2\\
      MD & 20-60-20 & 100M & 3.3 & 26.5\\
      LD & 20-20-60 & 100M & 7.5 & 60\\
    \hline
  \end{tabular}
  \caption{KV size distributions we use for our YCSB evaluation. Small KV pairs 
    are 33~B, medium KV pairs are 123~B, and large KV pairs are 1023~B.
  We report the record count, cache size per server, and dataset size used with 
  each KV size distribution.}
  \label{tab:kv-dist}
\end{table}

We examine the throughput (KOperations/s), efficiency (KCycles/operation), I/O
amplification, and network amplification of \name{} for the three
following setups: (1) without replication (No Replication), (2) with
replication, using our mechanism for sending the index to the
\backups{} (Send Index), and (3) with replication, where the
\backups{} perform compactions to build their index (Build Index),
which serves as a baseline. In Build Index, servers keeps additionally
an $L_0$ level in memory for their \backup{} regions, whereas in Send
Index, they do not. For these two setups two be equal, and since we
always use the same number of regions, in Build Index we configure
each region $L_0$ size to be half of the $L_0$ size used in the other
two setups.

We measure efficiency in cycles/op and define it as:
$$\mathvar{efficiency} = \frac{\frac{\mathvar{CPU\_utilization}}{100}
  \times \frac{cycles}{s} \times
  \mathvar{cores}}{\frac{\mathvar{average\_ops}}{s}} ~cycles/op,$$
where \mathvar{CPU\_utilization} is the average of CPU utilization
among all processors, excluding idle and I/O wait time, as given
by~\textit{mpstat}. As $cycles/s$ we use the per-core clock
frequency. $\mathvar{average\_ops}/s$ is the throughput reported by
YCSB, and $cores$ is the number of system cores including
hyperthreads.

I/O amplification measures the excess device traffic generated due to
compactions (for \primary{} and \backup{} regions) by \name{}, and we
define it as:
$$\mathvar{IO\_amplification} =
\frac{\mathvar{device\_traffic}}{\mathvar{dataset\_size}},$$ where
\mathvar{device\_traffic} is the total number of bytes read from or
written to the storage device and \mathvar{dataset\_size} is the total
size of all key-value requests issued during the experiment.

Lastly, network amplification is a measure of the excess network
traffic generated by \name{}, and we define it as:
$$\mathvar{network\_amplification} =
\frac{\mathvar{network\_traffic}}{\mathvar{dataset\_size}},$$ where
\mathvar{network\_traffic} is the total number of bytes sent by and
received from the servers' network cards.

\section{Experimental Evaluation}

In our evaluation of \name{} we answer the following questions:
\begin{enumerate}[itemsep=2pt]
  \item How does our \backup{} index construction (Send Index) method
    compare to performing compactions in \backup{} regions (Build
    Index) to construct the index?

  \item Where does \name{} spend its CPU cycles? How many cycles does
    Send Index save compared to Build Index for index construction?

  \item How does increasing the growth factor affect \name{}?

  \item Does Send Index improve performance and efficiency in
    small-dominated workloads, where KV separation gains diminish?
\end{enumerate}

\subsection{\name{} Performance and Efficiency}

\begin{figure}
  \includegraphics[width=\columnwidth]{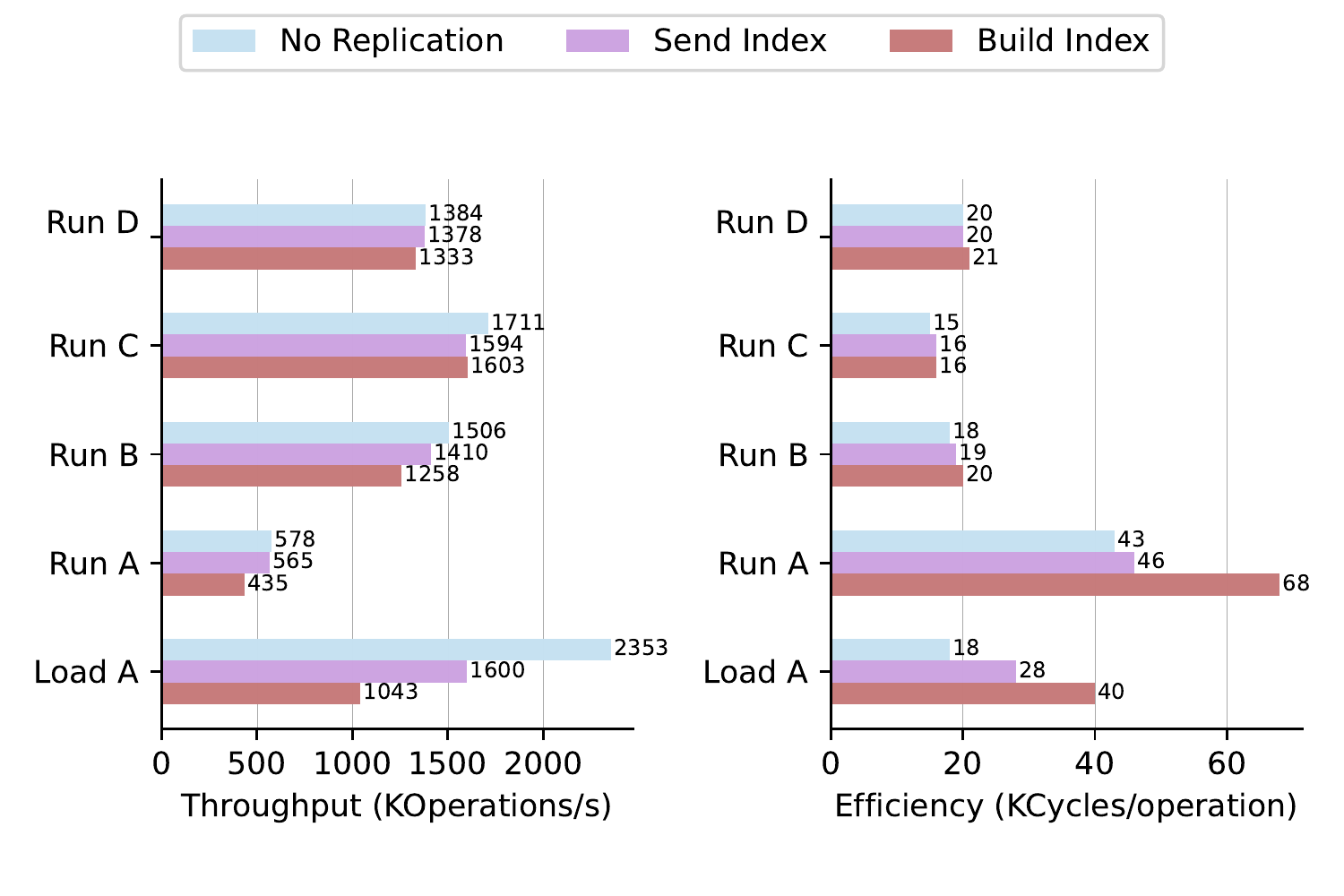}
  \caption{Performance and efficiency of \name{} for YCSB workloads
    Load A, Run A -- Run D with the SD KV size distribution.}
  \label{fig:a2d}
\end{figure}

In Figure~\ref{fig:a2d}, we evaluate \name{} using YCSB workloads Load
A and Run A -- Run D for the SD~\cite{fbook} workload. Since
replication doesn't impact read-dominated workloads, the performance
in workloads Run B -- Run D remains the same for all three
deployments. We focus the rest of our evaluation on the insert and
update heavy workloads Load A and Run A.

\begin{figure*}
  \centering \subfloat[width=\textwidth][Load A YCSB workload]{
    \includegraphics[width=\textwidth]{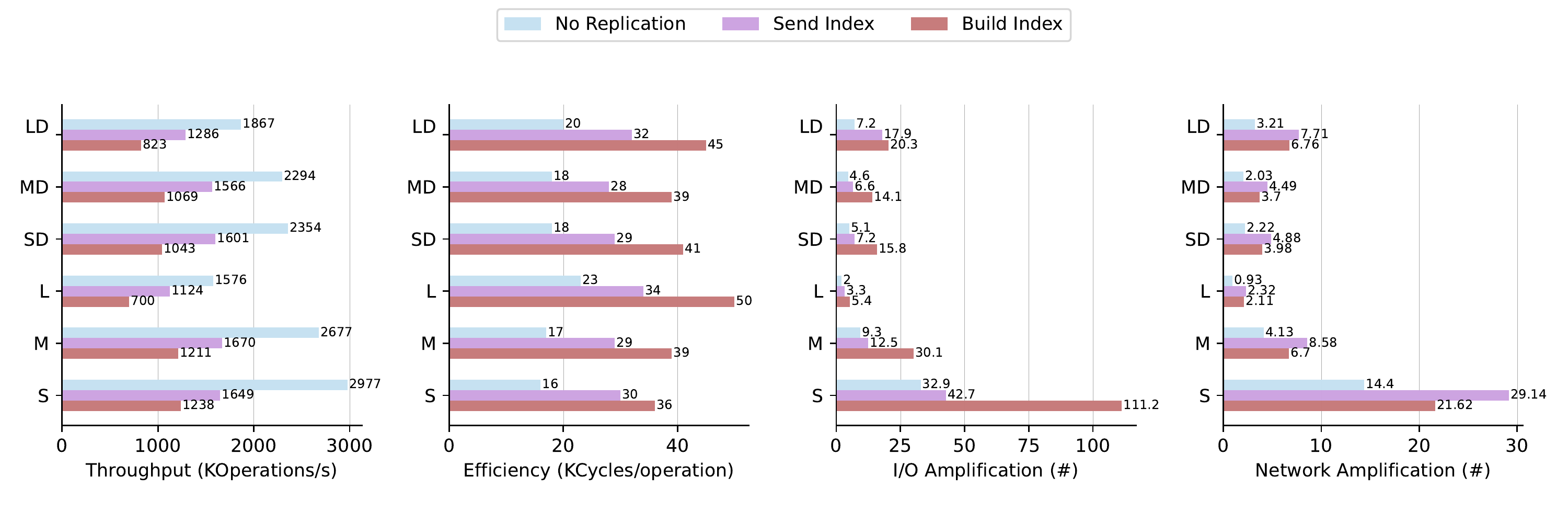}
    \label{fig:replication-loada}
  }

  \subfloat[width=\textwidth][Run A YCSB workload]{
    \includegraphics[width=\textwidth]{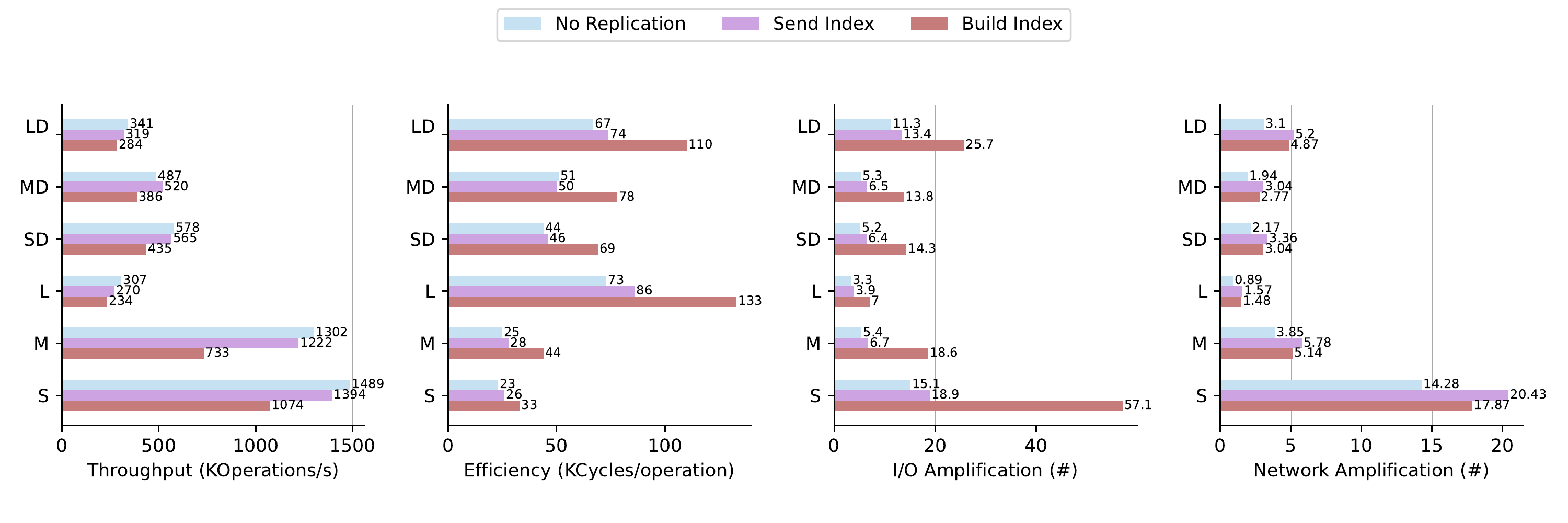}
    \label{fig:replication-runa}
  }
  \caption{Throughput, efficiency, I/O amplification, and network amplification 
    for the different key-value size distributions during the (a) YCSB Load A 
  and (b) Run A workloads.}
  \label{fig:replication}
\end{figure*}

\begin{figure}
  \includegraphics[width=\columnwidth]{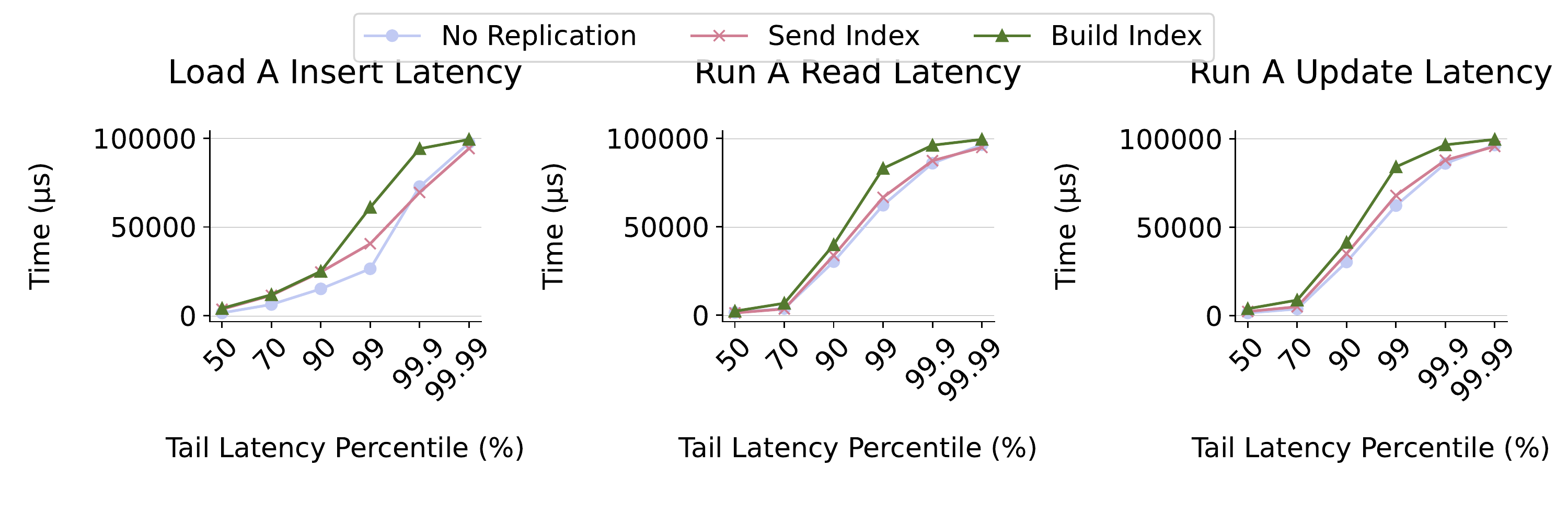}
  \caption{Tail latency for YCSB Load A and Run A workload operations
    using the SD key-value size distribution.}
  \label{fig:tail-latency-sd}
\end{figure}

We run Load A and Run A workloads for all six KV size distributions
and with growth factor 4 which minimizes I/O amplification (but not
space amplification). We set the $L_0$ size to 64K keys for the No Replication 
and Send Index configurations and to 32K keys for the Build Index 
configuration, since Build Index has twice as many $L_0$ indexes.
We measure throughput, efficiency, and I/O
amplification for the three different deployments explained in
Section~\ref{sec:platform}.  We summarize these results in
Figure~\ref{fig:replication}. We also report the tail latency in these
workloads for the SD KV size distribution in
Figure~\ref{fig:tail-latency-sd}.

Compared to Build Index, Send Index increases throughput by
$1.1-1.7\times$ for all KV size distributions, increases CPU
efficiency by $1.2-1.6\times$, and reduces I/O amplification by
$1.1-3.0\times$. Also, it is crucial to notice that compared to No
Replication, Build Index increases I/O amplification by $1.6 -
3.4\times$ while Send Index only increases I/O amplification by $1.4 -
1.5\times$, since eliminating compactions in \backup{} regions means
no additional read traffic for replication.
Furthermore, \name{} increases CPU efficiency by replacing expensive
I/O operations and key comparisons during compactions with a single
traversal of the new index segments and hash table accesses to rewrite
them.

Sending the \backup{} region indexes over the network increases
network traffic up to $1.2\times$. This trade-off favors \name{} since
it pays a slight increase in network traffic for increased efficiency
and decreased I/O amplification.

We also measure the tail latency for YCSB workloads Load A and Run A
using the SD KV size distribution. As shown in Figure
\ref{fig:tail-latency-sd}, Send Index improves the 99, 99.9, and
99.99\% tail latencies from $1.1$ to $1.5\times$ compared to Build
Index for all Load A and Run A operations.

\subsection{Cycles/Op Breakdown}

\begin{figure}
  \includegraphics[width=\columnwidth]{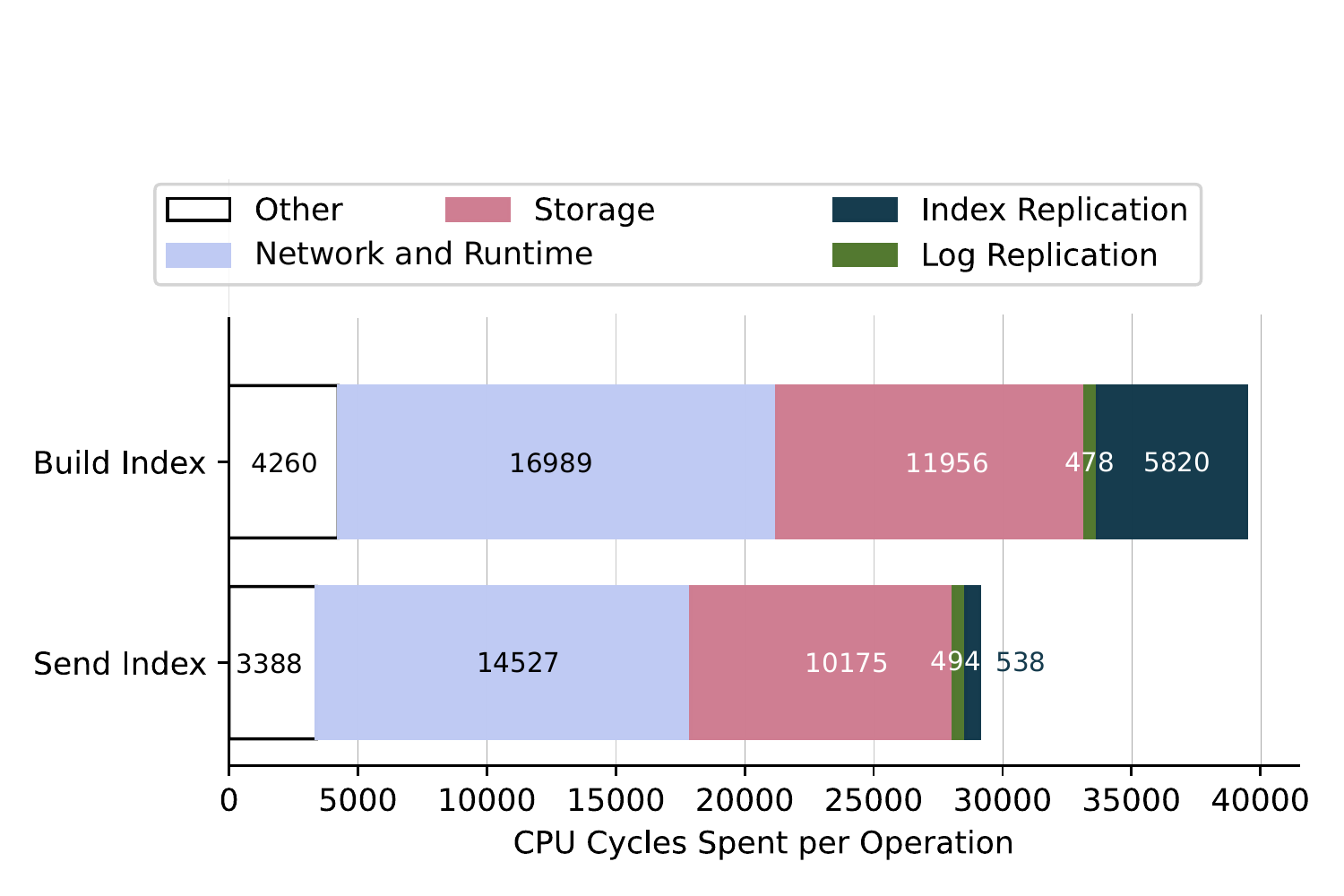}
  \caption{Breakdown of cycles spent per operation on network,
    storage, log replication and index replication.}
  \label{fig:breakdown}
\end{figure}

We run YCSB workloads Load A and Run A and profile \name{} using
\emph{perf} with call graph tracking enabled. We profile \name{} while
using Send Index and Build Index configurations.  We use the call
graph profiles generated to measure where CPU cycles are spent in
\name{} for Send Index and Build Index.  We count the CPU cycles spent
on four major parts of our system:

\begin{itemize}[itemsep=2pt]
  \item \textbf{Storage:} Cycles spent in KV store operations,
    excluding replication
  \item \textbf{Network and Runtime:} Cycles spent on detecting,
    scheduling, and processing client requests, excluding storage and
    replication
  \item \textbf{Log Replication:} Cycles spent on replicating KV pairs
    in a \backup{}'s value log
  \item \textbf{Index Replication:} Cycles spent to construct indexes
    for \backup{} regions. Send index spends these cycles to rewrite
    the index segments they receive from \primary{} \servers{}. Build
    Index spends these cycles on compactions and iterating KV value
    log segments to insert them into Kreon's $L_0$ index
  \item \textbf{Other:} All cycles not spent in the above categories
\end{itemize}

Figure~\ref{fig:breakdown} summarizes the results of our profiling.

\name{}'s Send Index technique requires 28\% fewer cycles than
performing compactions to construct \backup{} indexes. This 28\%
cycles amount to roughly 12K cycles per operation. They are divided
into: 5.5K fewer cycles for replicating \backup{} indexes, 2K fewer
cycles spent on storage, 2K fewer cycles spent on network and runtime
processing, and 2.5K fewer cycles spent on other parts of the
system. With the Send Index method, \name{} \servers{} spend
$10\times$ fewer cycles on constructing \backup{} indexes and
$1.36\times$ fewer cycles overall when compared to using the Build
Index method.

Sending the \primary{} index to \backups{} eliminates compactions for
backup regions resulting in increased CPU efficiency during \backup{}
index construction. While \backup{} \servers{} have to rewrite these
index segments, the rewriting process only involves hash table lookups
without requiring any read I/O, resulting in a more efficient
\backup{} index construction technique.

In comparison with Send Index, Build Index also spends 1.16x cycles on
network and runtime processing. This is due to additional queuing
effects which are a result of the increased load due to \backup{}
compactions.

\subsection{Impact of Growth Factor}

\begin{figure*}
  \includegraphics[width=\textwidth]{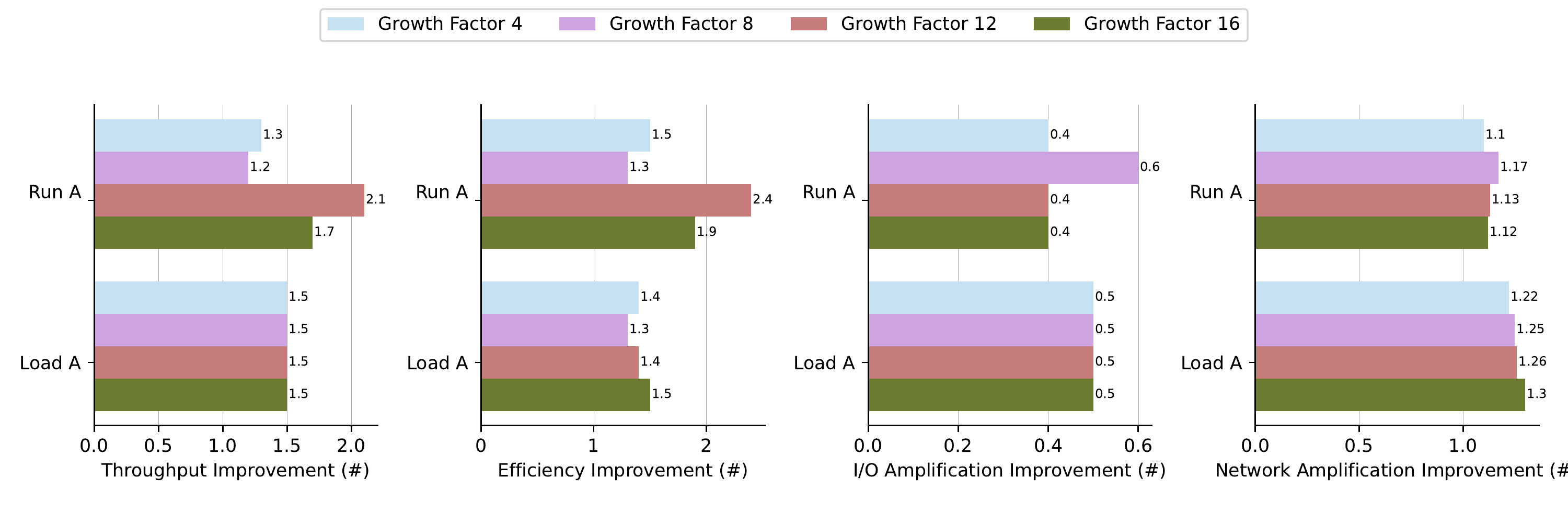}
  \caption{Send Index improvement over Build
    Index for Load A, Run A and different growth factors.}
  \label{fig:growth_factor}
\end{figure*}

In Figure \ref{fig:growth_factor}, we show that the gains in
performance, efficiency, and I/O amplification during Load A remain
constant when increasing the growth factor. However, during Run A, the
gains of our Send Index approach compared to Build Index
increase. Most notably, with growth factors 12 and 16, the performance
improvement is $2.1$ and $1.7\times$ respectively. Similarly,
efficiency is improved by $2.4$ and $1.9\times$, and I/O amplification
is decreased by $60\%$.

KV stores intentionally increase growth factor~\cite{vat,rocksdbspace}
(from 4 to 8-10) and pay the penalty of higher I/O amplification to
reduce space. In the Build Index, this penalty is further amplified to
two or three times according to the number of replicas per region.
However, Send Index eliminates these redundant compactions and allows
us to increase the growth factor and thus the space efficiency of LSM
tree-based KV stores without sacrificing significantly performance or
CPU efficiency.

\subsection{Small KVs Impact}

\begin{figure*}
  \centering \subfloat[width=\textwidth][Load A YCSB workload]{
    \includegraphics[width=\textwidth]{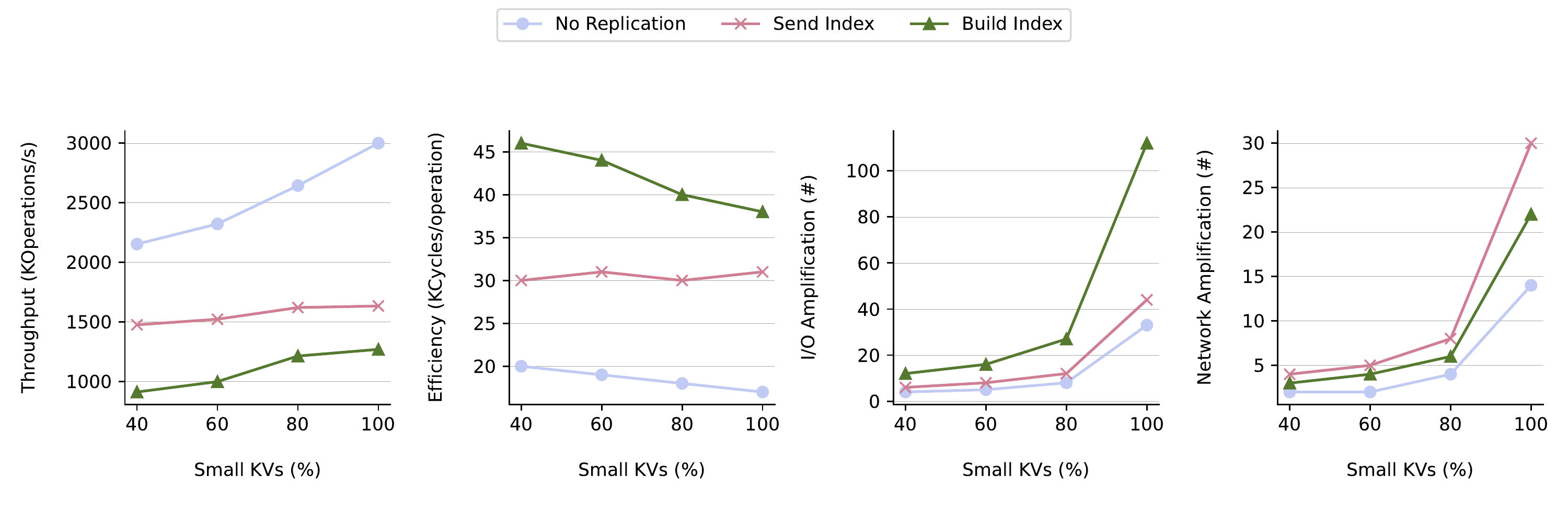}
    \label{fig:sd-loada}
  }

  \subfloat[width=\textwidth][Run A YCSB workload]{
    \includegraphics[width=\textwidth]{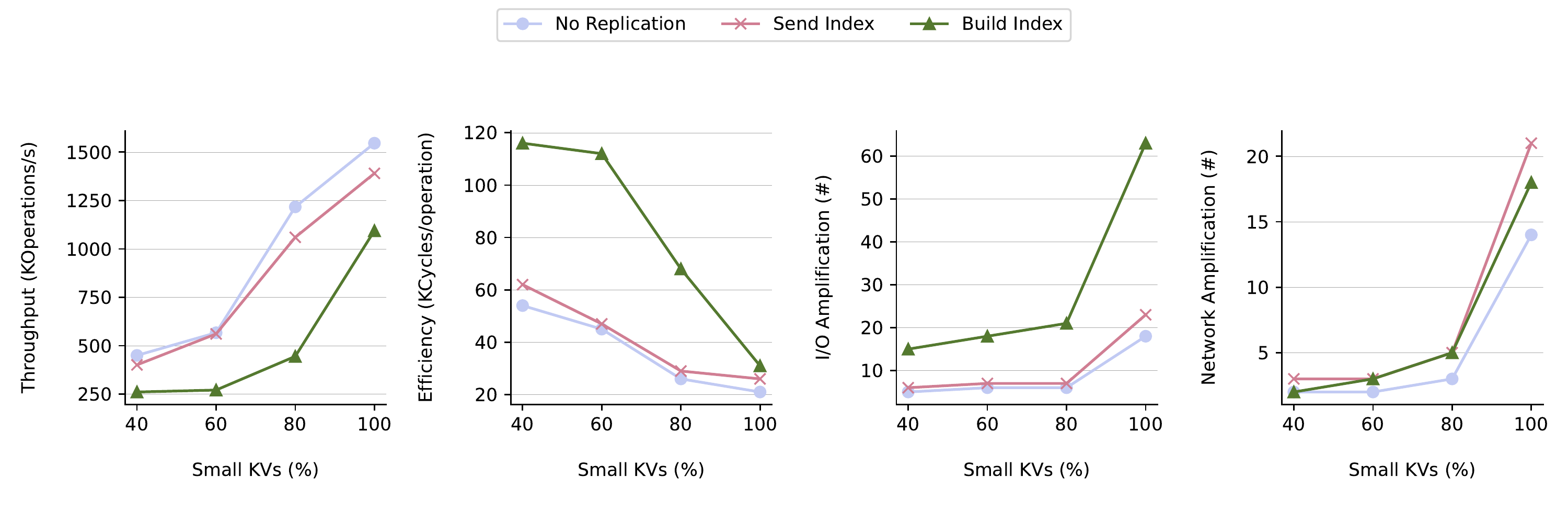}
    \label{fig:sd-runa}
  }
  \caption{Throughput, efficiency, I/O amplification, and network
    amplification for increasing percentages of small KVs during (a)
    YCSB Load A and (b) Run A workloads.}
  \label{fig:sd-var}
\end{figure*}

The KV separation~\cite{wisckey, kreon, hashkv} and hybrid placement
techniques~\cite{diffkv,parallax} gains in I/O amplification decrease
for $small \leq 33~B$ KV pairs, which are important for internet-scale
workloads~\cite{fbook}. This decrease is because the gains for KV
separation of small KV pairs is around
$2\times$~\cite{parallax}. However, if we include also the garbage
collection overheads, the gains further diminish making KV separation
identical to putting KVs in-place as RocksDB~\cite{rocksdb} does.

In this experiment, we investigate the impact that small KV pairs
percentage has on the efficiency of Send Index method. We set the
growth factor to 12 and examine four workloads where we vary small KV
pairs percentage to 40\%, 60\%, 80\%, and 100\%. In all four cases, we
equally divide the remaining percentage between medium and large KV
pairs.

As shown in Figure~\ref{fig:sd-var}, Send Index has from $1.2$ to
$2.3\times$ better throughput and efficiency than Build Index across
all workloads. I/O amplification for Build Index increases from $7.4$
to $9.3\times$.  From the above, we conclude that the Send Index
method has significant benefits even for workloads that consist of
80\%-90\% small KV pairs.

\section{Related Work}
\label{sec:related}

In this section we group related work in the following categories: (a)
LSM tree compaction offload techniques, (b) Log and index replication
techniques, and (c) efficient RDMA protocols for KV stores:

\paragraph{Compaction offload:}
Acazoo~\cite{acazoo} splits its data into shards and keeps replicas
for each shard using the ZAB~\cite{zookeeper} replication
protocol. Acazoo offloads compaction tasks from a shard's primary to
one of its replicas. Then, on compaction completion, it reconfigures the system 
through an election to make the server with the newly compacted data the 
primary.

Hailstorm~\cite{hailstorm} is a rack-scale
persistent KV store.  It uses a distributed filesystem to provide a
global namespace and scatters  SSTs across the
rack (in a deterministic way). Thus it can scale its I/O subsystem similar to
HBase/HDFS~\cite{hbase}. Unlike HBase, it schedules compaction tasks to
other servers through the global namespace offered by the distributed
filesystem substrate. Unlike these systems, \name{} can
efficiently keep both the primary and backup indexes up to date through the send 
index operation by using RDMA to perform efficient bulk network transfers.

\paragraph{Log and index replication techniques:}
Rose~\cite{rose} is a distributed replication engine that targets
racks with hard disk drives and TCP/IP networking where device I/O
is the bottleneck. In particular, it replicates data using a log
and builds the replica index by applying mutations in an LSM tree
index. The LSM tree removes random reads for updates and always
performs large I/Os. \name{} shares the idea of Rose to use the LSM tree
to build an index at the replica. However, it adapts its design for racks
that use fast storage devices and fast RDMA networks where the CPU is
the bottleneck. It does this by sending and rewriting the index and
removing redundant compactions at the \backups{}.

Tailwind~\cite{tailwind} is a replication protocol that uses RDMA
writes for data movement, whereas for control operations, it uses
conventional RPCs.  The primary server transfers log records to
buffers at the backup server by using one-sided RDMA writes.  Backup
servers are entirely passive; they flush their RDMA buffers to storage
periodically when the primary requests it. They have implemented and
evaluated their protocol on RAMCloud, a scale-out in-memory KV
store. Tailwind improves throughput and latency compared to
RAMCloud. \name{} adopts Tailwind's replication protocol
for its value log but further proposes a method to keep a backup index 
efficiently.


\paragraph{Efficient RDMA protocols for KV stores:}
Kalia \emph{et al.}~\cite{andersen} analyze different RDMA operations
and show that one-sided RDMA write operations provide the best
throughput and latency metrics. \name{} uses one-sided RDMA write
operations to build its protocol.

A second parameter is whether the KV store supports fixed or
variable size KVs. For instance, HERD~\cite{herd}, a
hash-based KV store, uses \emph{RDMA writes} to send
requests to the server, and \emph{RDMA send} messages to send a
reply back to the client. Send messages require a fixed maximum
size for KVs. \name{} uses only RDMA writes and
appropriate buffer management to support arbitrary KV sizes. HERD uses 
unreliable connections for RDMA writes, and an
unreliable datagram connection for RDMA sends. Note that they decide
to use RDMA send messages and unreliable datagram connections because RDMA write 
performance does not scale with the number of outbound connections in their 
implementation.  In addition, they show that
unreliable and reliable connections provide almost the same
performance. \name{} uses reliable connections
to reduce protocol complexity and examines their relative overhead
in persistent KV stores. We have not detected scalability
problems yet.

Other in-memory KV
stores~\cite{Mitchell:2013:UOR:2535461.2535475,Dragojevic:2014:FFR:2616448.2616486,Wei:2015:FIT:2815400.2815419}
use one-sided RDMA reads to offload read requests to the clients. \name{} does 
not use RDMA reads since lookups in LSM tree-based
systems are complex.  Typically, lookups and scan queries consist of multiple accesses to the devices
to fetch data. These data accesses must also be synchronized with compactions.

\section{Conclusions}
\label{sec:conc}

In this paper, we design \name{}, a replicated persistent LSM-based KV store
that targets racks with fast storage devices and fast network (RDMA).
\name{} implements an RDMA write-based client-server protocol and
replicates its data using an efficient RDMA write-based primary-backup
protocol.  \name{} takes a radical approach to keep an up-to-date
index at the \backups{} and avoid rebuilding it in case of a
failure. Instead of performing compactions at the \backup{} servers
(Build Index) \primary{} sends a pre-built index after each level
compaction (Send Index), trading a slight increase in network traffic for increased CPU
efficiency and decreased I/O amplification. \name{} implements an
efficient index rewrite mechanism at the \backups{}, which is used to translate
the \primary{} index's pointers into valid \backup{} index pointers. Compared
to Build Index, we
find that Send Index increases
throughput by up to $1.7\times$, CPU efficiency by up to $1.6\times$, decreases 
I/O amplification by up to $3.0\times$, and decreases tail latency by up to 
$1.5\times$ during YCSB Load A and Run A workloads.

Our approach enables KV stores
to operate with larger growth factors in order to
save space (6\% space saved by increasing the growth factor from 10 to 16),
without inducing
significant CPU overhead.  Furthermore, we show that Send Index provides 
significant benefits even in workloads where small KVs account for as much as 
90\% of the total operations.  We believe that the Send Index technique
can be adopted by state-of-the-art replicated LSM-based KV stores to
increase their CPU and space efficiency.

\bibliographystyle{abbrv}
\bibliography{paper}

\end{document}